\documentclass[aps,pra,nofootinbib,showpacs,twocolumn,superscriptaddress,10pt]{revtex4-1}

\usepackage{natbib}
\usepackage{amsmath,amssymb}
\usepackage{graphics}
\usepackage{color}

\begin{document}

\title{Biphoton states in correlated turbulence}

\author{Filippus S. \surname{Roux}}
\email{froux@nmisa.org}
\email{stef.roux@wits.ac.za}
\affiliation{National Metrology Institute of South Africa, Meiring Naud{\'e} Road, Brummeria, Pretoria, South Africa}
\affiliation{School of Physics, University of the Witwatersrand, Johannesburg 2000, South Africa}

\begin{abstract}
The effect of turbulence on a pair of photons propagating together through the same medium is analyzed. The behavior is compared to the case where these photons propagate separately through different turbulent media. The analysis is done with a multiple phase screen approach, by deriving and solving an infinitesimal propagation equation. We apply these results to the case where the initial photons are entangled in their spatial degrees of freedom with the aid of spontaneous parametric down-conversion. It is found that for this input state, the decay of entanglement in correlated media under the weak scintillation approximation is quicker than in uncorrelated media. Beyond the weak scintillation approximation, the entanglement in correlated media decays slower when it is close to zero --- approaching zero asymptotically as a function of scintillation strength. This is contrary to the case in uncorrelated media where entanglement becomes zero at a finite scintillation strength.
\end{abstract}


\maketitle

\section{Introduction}

While spatial modes, such as the orbital angular momentum (OAM) states of photons, allow high-dimensional free-space quantum communication, with the associated advantages of higher information capacity \cite{walborn} and increased security in quantum cryptography \cite{bp2000}, the distortion of these spatial modes, caused by turbulence in the atmosphere, adversely affects the performance of such a free-space quantum communication channel. For high-dimensional quantum key distribution protocols \cite{zeil2006,mafu} based on quantum entanglement, for instance, this distortion leads to a loss in quantum entanglement of the biphoton state \cite{konrad,tiersch}.

The decay of entanglement in biphoton states that are entangled in there spatial degrees of freedom has been studied theoretically \cite{sr,qturb4,qturb3,ipe, toddbrun,leonhard,notrunc}, numerically \cite{turbsim}, as well as experimentally \cite{pors,malik,oamturb,qkdturb}. There has also been a number of demonstrations of the use of OAM modes for the implementation of classical free-space communication links \cite{willner,krenn1} in addition to the free-space entanglement-based quantum key distribution, using OAM qubits \cite{vallone,krenn2}.

\begin{figure}[th]
\includegraphics{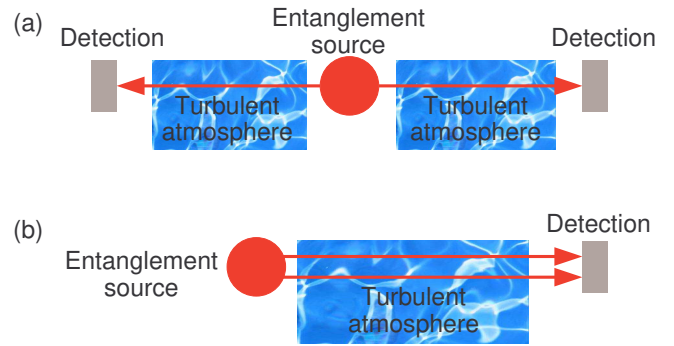}
\caption{Two different scenarios for two entangled photons propagating through turbulence. (a) The two photons propagate through different regions of turbulence. (b) The two photons propagate through the same turbulent medium.}
\label{scen}
\end{figure}

Usually it is assumed that the two entangled photons are sent through different uncorrelated regions of the turbulent medium [see Fig.~\ref{scen}(a)]. However, the portfolio of quantum technologies that are required for long distance quantum communication also includes quantum teleportation \cite{advtelport}. A recently proposed method to implement high-dimensional quantum teleportation \cite{teleoam} requires that multiple entangled photons are sent through the same channel. There are also other situations in which multiple photons would be sent through the same channel \cite{alonso}. As a result, in such scenarios, two or more photons that could be entangled, would see the same medium [see Fig.~\ref{scen}(b)]. In such a situation, the assumption of uncorrelated media is not valid anymore.

Here, we investigate the evolution of an entangled biphoton state when both photons propagate along the same path through a turbulent atmosphere. The analysis is based on the infinitesimal propagation approach \cite{ipe}. The latter is a multiple phase screen analysis, as opposed to the single phase screen analysis \cite{paterson}. Although most investigations into the evolution of photonic quantum states in turbulence employ a single phase screen analysis \cite{sr,qturb4,pors,malik,toddbrun,leonhard}, it is only valid under weak scintillation conditions \cite{turbsim}. The infinitesimal propagation approach, on the other hand, is valid under all conditions.

The current investigation follows the same approach to derive the required infinitesimal propagation equation (IPE), as used before \cite{notrunc}. However, the explicit derivation of the IPE for the continuous plane wave basis has not appeared in literature; previous derivations always assumed discrete bases \cite{ipe,lindb}. Implementations using such discrete bases tend to lead to truncation problems \cite{oamturb}. Therefore, we briefly show the derivation of the IPE for a single photon in the plane wave basis in Sec.~\ref{agter}, which also alleviates the discussion of the derivation required for correlated media and which is shown in Sec.~\ref{corr}. The resulting IPE for correlated media is solved, under the quadratic structure function approximation \cite{leader}, in Sec.~\ref{oplos}, leading to an integral expression for the evolving density matrix in terms of the input density matrix and a kernel function. For an illustration of the method, it is applied in Sec.~\ref{spdc} to the case where the input state is prepared with the aid of spontaneous parametric down-conversion (SPDC). Conclusions are given in Sec.~\ref{concl}.

\section{Background}
\label{agter}

Here we briefly review the basic principles of the derivation of the IPE. The idea is to consider the change in the density operator for a state propagating through an infinitesimally thin slab of a turbulent medium. This infinitesimal propagation is represented by an operator $dU$ such that
\begin{equation}
\hat{\rho}(z) \rightarrow \hat{\rho}(z+dz) = dU \hat{\rho}(z) dU^{\dag} ,
\label{infquop}
\end{equation}
where $\hat{\rho}(z)$ is the density operator for the quantum state as a function of the propagation distance $z$.

The effect of the operation on the density operator is readily expressed in terms of the change that such a thin slab of the turbulent medium produces in the state. For this purpose, one can start from the paraxial wave equation in a turbulent medium \cite{tatarski0,scintbook}
\begin{equation}
\nabla_T^2 g({\bf r}) - i 2k\partial_z g({\bf r}) + 2k^2 \tilde{n}({\bf r}) g({\bf r}) = 0 ,
\label{eomturb}
\end{equation}
where $g({\bf r})$ is the scalar electric field; ${\bf r}$ is the three-dimensional position vector; $k$ is the wavenumber and $\tilde{n}({\bf r})$ is the fluctuating part of the refractive index ($n=1+\tilde{n}$). The first two terms in Eq.~(\ref{eomturb}) represent the usual paraxial wave equation and the last term gives the effect of the turbulent medium.

The transformation of the electric field for an infinitesimal propagation along $z$ can be expressed with the aid of Eq.~(\ref{eomturb}). For this purpose, we perform a two-dimensional Fourier transform with respect to the transverse coordinates ($x,y$). Then we convert the remaining $z$-derivative into a finite difference. The result then reads
\begin{eqnarray}
G({\bf a},z+dz) & = & G({\bf a},z) + i dz \pi \lambda |{\bf a}|^2 G({\bf a},z) \nonumber \\
& & - i dz k N({\bf a},z) \otimes G({\bf a},z) ,
\label{transgft}
\end{eqnarray}
where $G({\bf a},z)$ and $N({\bf a},z)$ are the two-dimensional transverse Fourier transforms of $g({\bf x})$ and $\tilde{n}({\bf x})$, respectively, ${\bf a}$ is the two-dimensional transverse spatial frequency vector and $\otimes$ represents the convolution process. In the classical context, $G({\bf a},z)$ is a $z$-dependent angular spectrum, while, in the quantum context, it is interpreted as a two-dimensional Fourier domain wave function that evolves in $z$. As such, one can use it to represent a pure single photon state as
\begin{equation}
|\psi\rangle = \int |{\bf a}\rangle\ G({\bf a},z)\ {\rm d}^2 a ,
\label{puretoes}
\end{equation}
where $|{\bf a}\rangle$ represents the two-dimensional plane wave basis states. The effect of the infinitesimal propagation operation on this state then gives
\begin{equation}
dU |\psi\rangle = \int |{\bf a}\rangle\ G({\bf a},z+dz)\ {\rm d}^2 a ,
\label{puredz}
\end{equation}
where $G({\bf a},z+dz)$ is given by Eq.~(\ref{transgft}).

A general single photon state, expressed as a density operator, is given by
\begin{equation}
\hat{\rho}(z) = \int |{\bf a}_1\rangle F({\bf a}_1,{\bf a}_2,z)\langle {\bf a}_2|\ {\rm d}^2 a_1\ {\rm d}^2 a_2 ,
\label{enkelz}
\end{equation}
where $F({\bf a}_1,{\bf a}_2,z)$ is the density `matrix' in the plane wave basis. For a pure state, the density matrix factorizes
\begin{equation}
F({\bf a}_1,{\bf a}_2,z) = G({\bf a}_1,z) G^*({\bf a}_2,z) .
\label{suiwer}
\end{equation}
For a biphoton state, the density operator is
\begin{eqnarray}
\hat{\rho}(z) & = & \int |{\bf a}_1\rangle |{\bf a}_3\rangle F({\bf a}_1,{\bf a}_2,{\bf a}_3,{\bf a}_4,z)\langle {\bf a}_2|\langle {\bf a}_4| \nonumber \\
& & \times {\rm d}^2 a_1\ {\rm d}^2 a_2\ {\rm d}^2 a_3\ {\rm d}^2 a_4 .
\label{bifoton}
\end{eqnarray}
Here, ${\bf a}_1$ and ${\bf a}_2$ are associated with one photon and ${\bf a}_3$ and ${\bf a}_4$ with the other photon. When the propagation process is applied to ${\bf a}_2$ or ${\bf a}_4$, one needs to use the complex conjugate of the process given in Eq.~(\ref{transgft}).

The infinitesimal transformation shown in Eq.~(\ref{transgft}) can be expressed as an integral operation. We convert Eq.~(\ref{transgft}) into an integration over an small distance in $z$
\begin{eqnarray}
G({\bf a},z) & = & \int_{z_0}^{z} \int {\cal P}({\bf a},{\bf a}',z_1) G({\bf a}',z_1)\ {\rm d}^2 a'\ {\rm d}z_1 \nonumber \\
 & & + G({\bf a},z_0) ,
\label{transint}
\end{eqnarray}
where
\begin{equation}
{\cal P}({\bf a},{\bf a}',z) = i \pi \lambda |{\bf a}|^2 \delta({\bf a}-{\bf a}') - i k N({\bf a}-{\bf a}',z) .
\label{operdef}
\end{equation}
Applying the propagation process of Eq.~(\ref{operdef}) and its complex conjugate to the expression of the density matrix for the single photon density matrix, we get
\begin{eqnarray}
F({\bf a}_1,{\bf a}_2,z) & = &  \int_{z_0}^{z} \int [ {\cal P}({\bf a}_1,{\bf a}',z_1) F({\bf a}',{\bf a}_2,z_1) \nonumber \\
& & + {\cal P}^*({\bf a}_2,{\bf a}',z_1) F({\bf a}_1,{\bf a}',z_1) ]\ {\rm d}^2 a'\ {\rm d}z_1 \nonumber \\
& & + F({\bf a}_1,{\bf a}_2,z_0) .
\label{traenk}
\end{eqnarray}
The resulting expression contains terms with at most one factor of $N({\bf a},z)$.

The fluctuations in the refractive index, given in the transverse Fourier domain by $N({\bf a},z)$, is a stochastic function representing a particular realization of the turbulent medium. Since we do not have detailed information about any such particular realization, we need to compute the ensemble average over all possible realizations. It is assumed that the fluctuations have a zero average value $\langle N({\bf a},z)\rangle=0$. Hence, if we were to compute the ensemble average of Eq.~(\ref{traenk}), all the terms that contain $N({\bf a},z)$ would be removed, leaving only the free-space propagation terms without the effect of the turbulent medium.

To see the effect of turbulence, we need terms that are second order in $N({\bf a},z)$. For this purpose, we regard the right-hand side of Eq.~(\ref{traenk}) as the first order term in an expansion. The second order term is obtained by substituting the equation back into itself. Since the fluctuations are small, $N({\bf a},z)$ serves as an `expansion parameter.' The result is
\begin{widetext}
\begin{eqnarray}
F({\bf a}_1,{\bf a}_2,z) & = & \int_{z_0}^{z} \int {\cal P}({\bf a}_1,{\bf u},z_1) F({\bf u},{\bf a}_2,z_0) + {\cal P}^*({\bf a}_2,{\bf u},z_1) F({\bf a}_1,{\bf u},z_0) \nonumber \\
& & + \int_{z_0}^{z_1} \int {\cal P}({\bf a}_1,{\bf u},z_1) {\cal P}({\bf u},{\bf v},z_2) F({\bf v},{\bf a}_2,z_2) + {\cal P}({\bf a}_1,{\bf u},z_1) {\cal P}^*({\bf a}_2,{\bf v},z_2) F({\bf u},{\bf v},z_2) \nonumber \\
& & + {\cal P}^*({\bf a}_2,{\bf u},z_1) {\cal P}({\bf a}_1,{\bf v},z_2) F({\bf v},{\bf u},z_2) + {\cal P}^*({\bf a}_2,{\bf u},z_1) {\cal P}^*({\bf u},{\bf v},z_2) F({\bf a}_1,{\bf v},z_2) \nonumber \\
& & \times {\rm d}^2 v\ {\rm d}z_2\ {\rm d}^2 u\ {\rm d}z_1 + F({\bf a}_1,{\bf a}_2,z_0) .
\label{tweeord}
\end{eqnarray}
When we evaluate the ensemble averages, all the terms that contain only one factor of $N({\bf a},z)$ are removed, leaving only the free-space terms and terms with two factors of $N({\bf a},z)$. Some free-space terms have one $z$-integral, leading to a factor of $z-z_0=dz$, while others have two $z$-integrals leading to a factor of $dz^2/2$. Since the latter is a second order small number, all such terms are neglected. In the end, we find that those terms with one ${\cal P}$ only produce free-space terms, while those that contain two ${\cal P}$'s only produce terms that are second order in $N({\bf a},z)$. The resulting expression thus reads
\begin{eqnarray}
F({\bf a}_1,{\bf a}_2,z) & = & F({\bf a}_1,{\bf a}_2,z_0) + i dz \pi \lambda \left( |{\bf a}_1|^2 - |{\bf a}_2|^2 \right) F({\bf a}_1,{\bf a}_2,z_0) \nonumber \\
& & - k^2 \int \int_{z_0}^{z} \int_{z_0}^{z_1} \langle N({\bf a}_1-{\bf u},z_1) N({\bf u}-{\bf v},z_2)\rangle F({\bf v},{\bf a}_2,z_2) - \langle N({\bf a}_1-{\bf u},z_1) N^*({\bf a}_2-{\bf v},z_2)\rangle F({\bf u},{\bf v},z_2) \nonumber \\
& & - \langle N^*({\bf a}_2-{\bf u},z_1) N({\bf a}_1-{\bf v},z_2)\rangle F({\bf v},{\bf u},z_2) + \langle N^*({\bf a}_2-{\bf u},z_1) N^*({\bf u}-{\bf v},z_2)\rangle F({\bf a}_1,{\bf v},z_2) \nonumber \\
& & \times {\rm d}z_2\ {\rm d}z_1\ {\rm d}^2 v\ {\rm d}^2 u .
\label{tweeord0}
\end{eqnarray}
\end{widetext}

The refractive index fluctuations are represented by real-valued functions, which implies that $N^*({\bf a},z) = N(-{\bf a},z_2)$. Therefore, one can convert $N^*$ into $N$ and visa verse, until all terms contain the same combination of $N^*$ and $N$. One can assume that the $z$-dependences in the density matrices can be replace by $z_0$. (This is readily confirmed by performing another back substitution.) Then only the $N$ and $N^*$ contain $z$-dependences that need to be integrated over.

We now compute the general expression for
\begin{equation}
{\cal T}({\bf u},{\bf v}) \equiv \int_{z_0}^{z} \int_{z_0}^{z_1} \langle N({\bf u},z_1) N^*({\bf v},z_2)\rangle\ {\rm d}z_2\ {\rm d}z_1 .
\label{deft}
\end{equation}
For this purpose we model the stochastic functions by
\begin{equation}
N({\bf a},z) = \Delta^{-3/2} \int \exp(-i 2\pi c z) \chi({\bf k}) \sqrt{\Phi_n({\bf k})}\ {\rm d}c ,
\label{ndef}
\end{equation}
where $\Delta$ is the correlation distance in the Fourier domain; $\chi({\bf k})$ is a normally distributed random function with a zero mean; $\Phi_n({\bf k})$ is the refractive index power spectral density; and ${\bf k}$ is the three-dimensional Fourier domain coordinates. The transverse part of ${\bf k}$ is related to the transverse spatial frequency vector ${\bf k}_T=2\pi{\bf a}$ and the $z$-component is related to $c$ by $k_z=2\pi c$. The real-valued refractive index fluctuations require that $\chi^*({\bf k}) = \chi(-{\bf k})$ and they are assumed to be delta-correlated
\begin{equation}
\langle \chi({\bf k}_1) \chi^*({\bf k}_2)\rangle = (2\pi \Delta)^3 \delta({\bf k}_1-{\bf k}_2) .
\label{delkor}
\end{equation}
Using these properties, one can show that
\begin{eqnarray}
{\cal T}({\bf u},{\bf v}) & = & \delta({\bf u}-{\bf v}) \int \int_{z_0}^{z} \int_{z_0}^{z_1} \exp[-i 2\pi c (z_1-z_2)] \nonumber \\
& & \Phi_n({\bf k})\ {\rm d}z_2\ {\rm d}z_1\ {\rm d}c .
\label{deftt}
\end{eqnarray}
When we evaluate the two $z$-integrations and drop the anti-symmetric part of the result [which would not contribute to the final result due to the symmetry of the power spectral density $\Phi_n(-{\bf k})=\Phi_n({\bf k})$], we obtain
\begin{equation}
{\cal T}({\bf u},{\bf v}) = \delta({\bf u}-{\bf v}) \int \frac{1-\cos(2\pi c dz)}{(2\pi c)^2} \Phi_n({\bf k})\ {\rm d}c .
\label{deftt0}
\end{equation}

At this point we impose the Markov approximation \cite{scintbook}, which assumes that the turbulent medium is uncorrelated along the propagation direction. The effect is that one can set the $z$-component of ${\bf k}$ in the argument of the power spectral density to 0. Thus, the power spectral density becomes independent of $c$ and can be pulled out of the integral. One can then evaluate the integral over $c$, giving the result
\begin{equation}
{\cal T}({\bf u},{\bf v}) = \frac{dz}{2} \delta({\bf u}-{\bf v}) \Phi_0({\bf u}) ,
\label{deftta}
\end{equation}
where $\Phi_0({\bf u})\equiv\Phi_n(2\pi{\bf u},0)$.

Applying Eq.~(\ref{deftta}) in Eq.~(\ref{tweeord0}) and making a few simplifications, we obtain the expression for a single photon state
\begin{eqnarray}
\partial_z F({\bf a}_1,{\bf a}_2,z) & = & i \pi \lambda \left( |{\bf a}_1|^2 - |{\bf a}_2|^2 \right) F({\bf a}_1,{\bf a}_2,z) \nonumber \\
& & - k^2 \int \Phi_0({\bf u}) \left[ F({\bf a}_1,{\bf a}_2,z) \right. \nonumber \\
& & \left. - F({\bf a}_1-{\bf u},{\bf a}_2-{\bf u},z) \right]\ {\rm d}^2 u .
\label{enkipe}
\end{eqnarray}
Here we have converted the equation back into a differential equation in $z$. The resulting differential equation is the IPE for a single photon state.

\section{IPE in correlated media}
\label{corr}

Having reviewed the basic steps of the derivation of the single photon IPE, we next perform the derivation for the case where a biphoton propagates through the same medium. Note that, in the case where the two photons propagate through different media, one would have two stochastic functions $N_1$ and $N_2$ that are mutually uncorrelated so that $\langle N_1({\bf u},z_1) N_2^*({\bf v},z_2)\rangle=0$. The result is that the two photons act independently, leading to an IPE which is simply the duplicated version of the single photon IPE. When the two photons propagate through the same medium, there is only one stochastic function $N$. Therefore, additional terms appear due to the fact that the medium seen by one photon is perfectly correlated with the medium seen by the other photon. The resulting IPE is therefore more complicated.

We start by applying the propagation operation ${\cal P}$, given in Eq.~(\ref{operdef}), on the biphoton density matrix
\begin{eqnarray}
& & F({\bf a}_1,{\bf a}_2,{\bf a}_3,{\bf a}_4,z) \nonumber \\ & = &  \int_{z_0}^{z} \int [ {\cal P}({\bf a}_1,{\bf a}',z_1) F({\bf a}',{\bf a}_2,{\bf a}_3,{\bf a}_4,z_1) \nonumber \\
& & + {\cal P}^*({\bf a}_2,{\bf a}',z_1) F({\bf a}_1,{\bf a}',{\bf a}_3,{\bf a}_4,z_1) \nonumber \\
& & + {\cal P}({\bf a}_3,{\bf a}',z_1) F({\bf a}_1,{\bf a}_2,{\bf a}',{\bf a}_4,z_1) \nonumber \\
& & + {\cal P}^*({\bf a}_4,{\bf a}',z_1) F({\bf a}_1,{\bf a}_2,{\bf a}_3,{\bf a}',z_1) ]\ {\rm d}^2 a'\ {\rm d}z_1 \nonumber \\
& & + F({\bf a}_1,{\bf a}_2,{\bf a}_3,{\bf a}_4,z_0) .
\label{trabi}
\end{eqnarray}
Again, one needs to substitute this expression back into itself to produce a second order expansion in $N$. The result is the equivalent of Eq.~(\ref{tweeord}) for the biphoton case. Evaluating the ensemble averages, we again remove those terms with only one factor of $N$ or $N^*$. The remaining terms include the free-space terms for both photons and the dissipative terms, each with two factors of $N$ and/or $N^*$. However, since we allow both photons to propagate through the same medium, the dissipative terms are not only those that we found in the single photon case Eq.~(\ref{tweeord0}), duplicated for both photons, but also terms that involve both photons. The resulting expression is the equivalent of Eq.~(\ref{tweeord0}), but it contains 16 terms under the integral instead of just four.
Following the same steps as was done for the single photon, we reduce the 16 terms to seven terms and finally arrive at an IPE for correlated media given by
\begin{eqnarray}
\partial_z F & = & i\pi \lambda \left( |{\bf a}_1|^2 - |{\bf a}_2|^2 + |{\bf a}_3|^2 - |{\bf a}_4|^2 \right) F \nonumber \\
& & - k^2 \int \Phi_0({\bf u}) \left[ 2 F({\bf a}_1,{\bf a}_2,{\bf a}_3,{\bf a}_4,z) \right. \nonumber \\
& & - F({\bf a}_1-{\bf u},{\bf a}_2-{\bf u},{\bf a}_3,{\bf a}_4,z) \nonumber \\
& & - F({\bf a}_1,{\bf a}_2,{\bf a}_3-{\bf u},{\bf a}_4-{\bf u},z) \nonumber \\
& & - \xi F({\bf a}_1-{\bf u},{\bf a}_2,{\bf a}_3,{\bf a}_4-{\bf u},z) \nonumber \\
& & - \xi F({\bf a}_1,{\bf a}_2-{\bf u},{\bf a}_3-{\bf u},{\bf a}_4,z) \nonumber \\
& & + \xi F({\bf a}_1-{\bf u},{\bf a}_2,{\bf a}_3+{\bf u},{\bf a}_4,z) \nonumber \\
& & \left. + \xi F({\bf a}_1,{\bf a}_2-{\bf u},{\bf a}_3,{\bf a}_4+{\bf u},z) \right]\ {\rm d}^2 u .
\label{ipebi}
\end{eqnarray}
The first three of the seven terms under the integral are the same terms one would obtain for the case where the two photons propagate through separate uncorrelated media. They represent a duplication of the two terms obtained for the single photon case, shown in Eq.~(\ref{enkipe}). The last four terms represent the correlation terms that appear because the two photons are propagating through the same medium. To keep track of these correlation terms, we label them with a tag $\xi$. For $\xi=0$ we'll recover the uncorrelated case and for $\xi=1$ we have the correlated case.

\section{Position domain equation}
\label{pos}

The expression of the IPE in Eq.~(\ref{ipebi}) contains the density matrix to be solved under an integral. This makes it difficult to solve the equation directly in the given form. To enable one to solve the equation, it needs to be converted to a different form that separates the density matrix from the integral. We do this by expressing the density matrix in the equation in terms of a Fourier transform
\begin{eqnarray}
F({\bf a}_1,{\bf a}_2,{\bf a}_3,{\bf a}_4,z) & = & \int \exp[i2\pi ({\bf a}_1\cdot{\bf x}_1 - {\bf a}_2\cdot{\bf x}_2 \nonumber \\
& & + {\bf a}_3\cdot{\bf x}_3 - {\bf a}_4\cdot{\bf x}_4)] \nonumber \\
& & \times f({\bf x}_1,{\bf x}_2,{\bf x}_3,{\bf x}_4,z) \nonumber \\
& & \times {\rm d}^2 x_1\ {\rm d}^2 x_2\ {\rm d}^2 x_3\ {\rm d}^2 x_4 ,
\end{eqnarray}
and then evaluate the inverse Fourier transform of the entire expression. The free-space propagation terms become second-order spatial derivatives, with respect to all transverse coordinates
\begin{equation}
\partial_z f = \frac{-i}{2k} \left( \nabla_1^2 - \nabla_2^2 + \nabla_3^2 - \nabla_4^2 \right) f - k^2 Q f ,
\label{dvf}
\end{equation}
where
\begin{equation}
\nabla_n^2=\frac{\partial^2}{\partial x_n} + \frac{\partial^2}{\partial y_n} ,
\label{lapdef}
\end{equation}
with $n=\{1,2,3,4\}$, and $Q$ is given by
\begin{eqnarray}
Q & = & \int \Phi_0({\bf u}) \left\{ 2  - \cos[2\pi {\bf u}\cdot({\bf x}_1 - {\bf x}_2)] \right. \nonumber \\
& & - \cos[2\pi {\bf u}\cdot({\bf x}_3 - {\bf x}_4)] - \xi \cos[2\pi {\bf u}\cdot({\bf x}_3 - {\bf x}_2)] \nonumber \\
& & - \xi \cos[2\pi {\bf u}\cdot({\bf x}_1 - {\bf x}_4)] + \xi \cos[2\pi {\bf u}\cdot({\bf x}_1 - {\bf x}_3)] \nonumber \\
& & \left. + \xi \cos[2\pi {\bf u}\cdot({\bf x}_4 - {\bf x}_2)] \right\}\ {\rm d}^2 u ,
\label{qdefc}
\end{eqnarray}
where we used the symmetry of $\Phi_0({\bf u})$.

The integration of $Q$ can be evaluated for a given expression of the power spectral density. Using, for this purpose, the Kolmogorov power spectral density \cite{scintbook}
\begin{equation}
\Phi_n ({\bf k}) = 0.033 (2\pi)^3 C_n^2 |{\bf k}|^{-11/3} ,
\label{klmgrv}
\end{equation}
where $C_n^2$ is the refractive index structure constant, we find that
\begin{equation}
\int \Phi_0({\bf u}) \cos(2\pi {\bf u}\cdot{\bf x})\ {\rm d}^2 u = \Lambda_0 - {\cal S} C_n^2|{\bf x}|^{5/3} ,
\label{Qx}
\end{equation}
where ${\cal S}=1.457$ and
\begin{equation}
\Lambda_0 = \int \Phi_0({\bf u})\ {\rm d}^2 u
\label{verw7}
\end{equation}
is a divergent quantity (in the limit of infinite outer scale). Fortunate, $\Lambda_0$ cancels out in the final expression for $Q$, which reads
\begin{eqnarray}
Q & = & {\cal S} C_n^2 \left( |{\bf x}_1 - {\bf x}_2|^{5/3} + |{\bf x}_3 - {\bf x}_4|^{5/3} \right. \nonumber \\
& & + \xi |{\bf x}_3 - {\bf x}_2|^{5/3} + \xi |{\bf x}_1 - {\bf x}_4|^{5/3} \nonumber \\
& & \left. - \xi |{\bf x}_1 - {\bf x}_3|^{5/3} - \xi |{\bf x}_4 - {\bf x}_2|^{5/3} \right) .
\label{qopl}
\end{eqnarray}

The powers of $5/3$ in Eq.~(\ref{qopl}) makes the solution of Eq.~(\ref{dvf}) challenging. For this reason, we employ the quadratic structure function approximation \cite{leader} and replace $5/3\rightarrow 2$. We also compensate for the change in the dimension of the expression by inserting a factor of the transverse scale with an appropriate power. For the transverse scale we use the radius of the optical beam $w_0$. Thus we obtain
\begin{eqnarray}
Q & = & \zeta \left[ |{\bf x}_1 - {\bf x}_2|^2 + |{\bf x}_3 - {\bf x}_4|^2 \right. \nonumber \\
& & \left. + 2 \xi ({\bf x}_1 - {\bf x}_2)\cdot( {\bf x}_3 - {\bf x}_4) \right] ,
\label{qkopl}
\end{eqnarray}
where we defined
\begin{equation}
\zeta = \frac{{\cal S} C_n^2}{w_0^{1/3}} ,
\label{kappadef}
\end{equation}
for the sake of having cleaner expressions. Eventually the latter will be incorporated into dimensionless combinations of the dimension parameters.

\section{Solution}
\label{oplos}

To find a solution for the differential equation in Eq.~(\ref{dvf}), we need to follow several steps, involving partial solutions that are obtained by removing all terms that contain derivatives with respect to transverse coordinates. At some stages, the resulting equation only contains terms consisting of such derivatives. Then one performs a Fourier transform with respect to these coordinates to remove the derivatives.

These steps work better when they are done in terms of sums and differences of the coordinates. For this reason, as a first step, we convert the expression into such sums and differences, using the definitions
\begin{equation}
\begin{aligned}
{\bf x}_{s1} & = \frac{1}{2}({\bf x}_1+{\bf x}_2) \\
{\bf x}_{d1} & = {\bf x}_1-{\bf x}_2 \\
{\bf x}_{s2} & = \frac{1}{2}({\bf x}_3+{\bf x}_4) \\
{\bf x}_{d2} & = {\bf x}_3-{\bf x}_4 .
\end{aligned}
\end{equation}
The differential equation then becomes
\begin{equation}
\partial_z h = \frac{-i}{k} \left( \nabla_{s1} \cdot \nabla_{d1} + \nabla_{s2} \cdot \nabla_{d2} \right) h - k^2 Q h ,
\label{dvh}
\end{equation}
where $h=h({\bf x}_{s1},{\bf x}_{d1},{\bf x}_{s2},{\bf x}_{d2},z)$ is the density matrix expressed in terms of the sum- and difference-coordinates,
\begin{equation}
\nabla_{n} = \hat{x} \frac{\partial}{\partial x_{n}} + \hat{y} \frac{\partial}{\partial y_{n}} ,
\end{equation}
with $n=\{s1,d1,s2,d2\}$, and
\begin{equation}
Q = \zeta \left( |{\bf x}_{d1}|^2 + |{\bf x}_{d2}|^2 + 2 \xi {\bf x}_{d1} \cdot {\bf x}_{d2} \right) .
\label{qdefsv}
\end{equation}

Next, we use a partial solution that removes the last term in Eq.~(\ref{dvh}). For this purpose we use the anzats
\begin{equation}
h = h_1 \exp(- k^2 z Q ) ,
\label{popl1}
\end{equation}
where $h_1=h_1({\bf x}_{s1},{\bf x}_{d1},{\bf x}_{s2},{\bf x}_{d2},z)$ is a new density matrix, still to be solved. By substituting Eq.~(\ref{popl1}) into Eq.~(\ref{dvh}), we derive a differential equation for $h_1$, given by
\begin{eqnarray}
\partial_z h_1 & = & \frac{-i}{k} \left( \nabla_{s1} \cdot \nabla_{d1} + \nabla_{s2} \cdot \nabla_{d2} \right) h_1 \nonumber \\
& & + i 2 \zeta k z \left[ ({\bf x}_{d1}+\xi {\bf x}_{d2}) \cdot \nabla_{s1} \right. \nonumber \\
& & \left. + (\xi {\bf x}_{d1}+{\bf x}_{d2}) \cdot \nabla_{s2} \right] h_1 .
\label{dvh1}
\end{eqnarray}
Note that $\xi$, which tags the correlation terms, governs the structure of the equation.

All the terms now represent derivatives of $h_1$. We perform a Fourier transform in the sum coordinates to remove some of these derivatives
\begin{eqnarray}
& & h_1({\bf x}_{s1},{\bf x}_{d1},{\bf x}_{s2},{\bf x}_{d2},z) \nonumber \\ & = & \int H_1({\bf a}_{d1},{\bf x}_{d1},{\bf a}_{d2},{\bf x}_{d2},z) \nonumber \\
& & \times \exp[-i 2\pi({\bf x}_{s1}\cdot{\bf a}_{d1}+{\bf x}_{s2}\cdot{\bf a}_{d2})] \nonumber \\
& & \times {\rm d}^2 a_{d1}\ {\rm d}^2 a_{d2} .
\label{ft1}
\end{eqnarray}
The resulting differential equation for $H_1$ is
\begin{eqnarray}
\partial_z H_1 & = & -\lambda \left( {\bf a}_{d1}\cdot \nabla_{d1} + {\bf a}_{d2}\cdot \nabla_{d2} \right) H_1 \nonumber \\
& & + 4\pi\zeta k z \left( {\bf x}_{d1}\cdot{\bf a}_{d1}+{\bf x}_{d2}\cdot{\bf a}_{d2} \right. \nonumber \\
& & \left. + \xi {\bf x}_{d1}\cdot{\bf a}_{d2} + \xi {\bf x}_{d2}\cdot{\bf a}_{d1} \right) H_1 .
\label{dvH1}
\end{eqnarray}

We proceed, as before, by constructing partial solutions that remove the non-derivative terms. In this case, we do it twice in a row. First, we have the anzats
\begin{eqnarray}
H_1 & = & H_2 \exp\left[ 2\pi\zeta k z^2 \left( {\bf x}_{d1}\cdot{\bf a}_{d1}+{\bf x}_{d2}\cdot{\bf a}_{d2} \right. \right. \nonumber \\
& & \left. \left. + \xi {\bf x}_{d1}\cdot{\bf a}_{d2} + \xi {\bf x}_{d2}\cdot{\bf a}_{d1} \right) \right] ,
\label{popl2}
\end{eqnarray}
leading to
\begin{eqnarray}
\partial_z H_2 & = & - 4\pi^2\zeta z^2 \left( |{\bf a}_{d1}|^2+|{\bf a}_{d2}|^2 + 2 \xi {\bf a}_{d1}\cdot{\bf a}_{d2} \right) H_2 \nonumber \\
& & -\lambda \left( {\bf a}_{d1}\cdot \nabla_{d1} + {\bf a}_{d2}\cdot \nabla_{d2} \right) H_2 .
\label{dvH2}
\end{eqnarray}
Then we use the anzats
\begin{equation}
H_2 = H_3 \exp\left[ - \frac{4\pi^2}{3} \zeta z^3 \left( |{\bf a}_{d1}|^2+|{\bf a}_{d2}|^2 + 2 \xi {\bf a}_{d1}\cdot{\bf a}_{d2} \right) \right]
\label{popl3}
\end{equation}
which leads to
\begin{equation}
\partial_z H_3 = -\lambda \left( {\bf a}_{d1}\cdot \nabla_{d1} + {\bf a}_{d2}\cdot \nabla_{d2} \right) H_3 .
\label{dvH3}
\end{equation}

Again, we reach a point where all the remaining terms are derivatives. As before, we remove them with a Fourier transform; this time, with respect to all the difference coordinates
\begin{eqnarray}
& & H_3({\bf a}_{d1},{\bf x}_{d1},{\bf a}_{d2},{\bf x}_{d2},z) \nonumber \\ & = & \int L_1({\bf a}_{d1},{\bf a}_{s1},{\bf a}_{d2},{\bf a}_{s2},z) \nonumber \\
& & \times \exp[-i 2\pi({\bf x}_{d1}\cdot{\bf a}_{s1}+{\bf x}_{d2}\cdot{\bf a}_{s2})] \nonumber \\
& & \times {\rm d}^2 a_{s1}\ {\rm d}^2 a_{s2} .
\label{ft2}
\end{eqnarray}

The differential equation for $L_1$ is given by
\begin{equation}
\partial_z L_1 = i2\pi\lambda \left( {\bf a}_{d1}\cdot{\bf a}_{s1}+{\bf a}_{d2}\cdot{\bf a}_{s2} \right) L_1
\label{dvL1}
\end{equation}
and now it has a full solution, given by
\begin{eqnarray}
L_1 & = & L_0 \exp\left[ i2\pi\lambda z \left( {\bf a}_{d1}\cdot{\bf a}_{s1}+{\bf a}_{d2}\cdot{\bf a}_{s2} \right) \right] ,
\label{popl4}
\end{eqnarray}
where $L_0=L_0({\bf a}_{d1},{\bf a}_{s1},{\bf a}_{d2},{\bf a}_{s2})$ is the initial density matrix at $z=0$.

The complete solution is obtained by substituting Eqs.~(\ref{popl4}), (\ref{ft2}), (\ref{popl3}), (\ref{popl2}), (\ref{ft1}) and (\ref{popl1}) consecutively back into each other. Thus we obtain
\begin{widetext}
\begin{eqnarray}
h({\bf x}_{s1},{\bf x}_{d1},{\bf x}_{s2},{\bf x}_{d2},z) & = & \int \int L_0({\bf a}_{d1},{\bf a}_{s1},{\bf a}_{d2},{\bf a}_{s2}) \exp\left[ i2\pi\lambda z \left( {\bf a}_{d1}\cdot{\bf a}_{s1}+{\bf a}_{d2}\cdot{\bf a}_{s2} \right) \right] \nonumber \\
& & \times \exp\left[ 2\pi\zeta k z^2 \left( {\bf x}_{d1}\cdot{\bf a}_{d1}+{\bf x}_{d2}\cdot{\bf a}_{d2} + \xi {\bf x}_{d1}\cdot{\bf a}_{d2} + \xi {\bf x}_{d2}\cdot{\bf a}_{d1} \right) \right] \nonumber \\
& & \times \exp\left[ - \frac{4\pi^2}{3} \zeta z^3 \left( |{\bf a}_{d1}|^2+|{\bf a}_{d2}|^2 + 2 \xi {\bf a}_{d1}\cdot{\bf a}_{d2} \right) \right] \nonumber \\
& & \times \exp[-i 2\pi({\bf x}_{s1}\cdot{\bf a}_{d1}+{\bf x}_{s2}\cdot{\bf a}_{d2}+{\bf x}_{d1}\cdot{\bf a}_{s1}+{\bf x}_{d2}\cdot{\bf a}_{s2})] \nonumber \\
& & \times \exp\left[- k^2 z \zeta \left( |{\bf x}_{d1}|^2 + |{\bf x}_{d2}|^2 + 2 \xi {\bf x}_{d1} \cdot {\bf x}_{d2} \right) \right] \nonumber \\
& & \times {\rm d}^2 a_{d1}\ {\rm d}^2 a_{d2}\ {\rm d}^2 a_{s1}\ {\rm d}^2 a_{s2} .
\label{volopl}
\end{eqnarray}
The resulting expression relates an initial density matrix in terms of sums and differences in the Fourier domain coordinates to the final density matrix in terms of sums and differences in position domain coordinates. It is convenient to work with the expressions in the Fourier domain. Therefore, we perform a Fourier transform on the expression of the complete solution in Eq.~(\ref{volopl}). However, it is necessary at this point to select the particular case by either setting $\xi=1$ for propagation through the same correlated medium or setting $\xi=0$ for propagation through different uncorrelated media. Different expressions are obtained for the two cases. We also convert the coordinates back to their original form by undoing the sums and differences.

In the case of propagation through the same correlated medium ($\xi=1$), we obtain
\begin{eqnarray}
R({\bf a}_1,{\bf a}_2,{\bf a}_3,{\bf a}_4,t) & = & \frac{\pi w_0^2}{2 {\cal K} t} \exp\left[ i\pi^2 w_0^2 t \left( |{\bf a}_1|^2-|{\bf a}_2|^2+|{\bf a}_3|^2-|{\bf a}_4|^2 \right) \right] \nonumber \\
& & \times \int R_0({\bf a}_1-{\bf u},{\bf a}_2-{\bf u},{\bf a}_3-{\bf u},{\bf a}_4-{\bf u}) \exp\left[ -i\pi^2 w_0^2 t \left( {\bf a}_1-{\bf a}_2+{\bf a}_3-{\bf a}_4 \right)\cdot{\bf u} \right] \nonumber \\
& & \times \exp\left( -\frac{\pi^2}{6} w_0^2 {\cal K} t^3 |{\bf a}_1-{\bf a}_2+{\bf a}_3-{\bf a}_4|^2 -\frac{\pi^2 w_0^2 |{\bf u}|^2}{2 {\cal K} t} \right)\ {\rm d}^2 u .
\label{koropl}
\end{eqnarray}
In the case of propagation through different uncorrelated media ($\xi=0$), the result reads
\begin{eqnarray}
R({\bf a}_1,{\bf a}_2,{\bf a}_3,{\bf a}_4,t) & = & \frac{\pi^2 w_0^4}{4 {\cal K}^2 t^2} \exp\left[ i\pi^2 w_0^2 t \left( |{\bf a}_1|^2-|{\bf a}_2|^2+|{\bf a}_3|^2-|{\bf a}_4|^2 \right) \right] \nonumber \\
& & \times \int R_0({\bf a}_1-{\bf u}_1,{\bf a}_2-{\bf u}_1,{\bf a}_3-{\bf u}_2,{\bf a}_4-{\bf u}_2)
\exp\left\{ -i\pi^2 w_0^2 t \left[ ({\bf a}_1-{\bf a}_2)\cdot{\bf u}_1+({\bf a}_3-{\bf a}_4)\cdot{\bf u}_2 \right] \right\} \nonumber \\
& & \times \exp\left[ -\frac{\pi^2}{6} w_0^2 {\cal K} t^3 \left( |{\bf a}_1-{\bf a}_2|^2 + |{\bf a}_3-{\bf a}_4|^2 \right) -\frac{\pi^2 w_0^2}{2 {\cal K} t} \left( |{\bf u}_1|^2 + |{\bf u}_2|^2 \right) \right]\ {\rm d}^2 u_1\ {\rm d}^2 u_2 .
\label{onkopl}
\end{eqnarray}
\end{widetext}
Here we defined a normalized propagation distance
\begin{equation}
t = \frac{z}{z_R} = \frac{z\lambda}{\pi w_0^2} ,
\label{tdef}
\end{equation}
and a dimensionless turbulence strength
\begin{equation}
{\cal K} = \frac{2\pi^3 {\cal S} C_n^2 w_0^{11/3}}{\lambda^3} .
\label{kdef}
\end{equation}
The main result of this paper is the expression for the density matrix of a biphoton propagating together through the same (correlated) medium, given in Eq.~(\ref{koropl}). The expression for propagation through uncorrelated media, given in Eq.~(\ref{onkopl}), agrees with what was obtained previously [see Eq.~(24) in Ref \cite{notrunc}, with a change in the sign of the integration variables]. Below, we consider an application that allows us to compare the correlated and uncorrelated cases. We also compare these results in the single phase screen approximation, which requires a brief discussion of the weak scintillation limit.

\section{Weak scintillation limit}
\label{weakscint}

In general, the Rytov variance, which is given by
\begin{equation}
\sigma_R^2 = 1.23 C_n^2 k^{7/6} z^{11/6} ,
\label{rytov}
\end{equation}
is considered to be a good indication of scintillation strength. The condition for weak scintillation is $\sigma_R^2 < C$, where $C$ is a constant of $\sim O(1)$ \cite{scintbook}.

The single phase screen approach \cite{paterson}, shows that the evolution of photonic states under weak scintillation only depends on a dimensionless combination of the dimension parameters, given by \cite{sr}
\begin{equation}
{\cal W} = \frac{w_0}{r_0} ,
\label{wdef}
\end{equation}
where
\begin{equation}
r_0 = 0.185 \left(\frac{\lambda^2}{C_n^2 z}\right)^{3/5} ,
\label{fried}
\end{equation}
is the Fried parameter \cite{fried}. If one expresses the Rytov variance in terms of ${\cal W}$ and ${\cal K}$, given in Eqs.~(\ref{wdef}) and (\ref{kdef}), respectively,  one obtains
\begin{equation}
\sigma^2_R = \frac{2.57 {\cal W}^{55/18}}{{\cal K}^{5/6}} .
\label{RytKW}
\end{equation}
It then follows that, for constant ${\cal W}$, the scintillation strength $\sigma^2_R$ would decrease to zero in the limit where ${\cal K}\rightarrow\infty$. In other words, weak scintillation requires strong turbulence. Since ${\cal W}$ also depends on the turbulence strength through $C_n^2$, one needs to take the limit $z\rightarrow 0$ at the same time, in such a way that ${\cal W}$ remains constant.

To apply the weak scintillation limit to the IPE results, one first needs to replace
\begin{equation}
t \rightarrow \frac{1.72 {\cal W}^{5/3}}{\cal K} .
\label{vervt}
\end{equation}
In the limit ${\cal K}\rightarrow\infty$, the IPE results then reproduce the single phase screen results. We computed the single phase screen results both through direct calculations and by applying this weak scintillation limit to our IPE results. The agreement in the expressions that we obtained provides an independent cross-check for our calculations.

\section{Application: SPDC state}
\label{spdc}

Here, we consider the situation where the biphoton states are prepared in an SPDC process. Such a state can be expressed in the Fourier domain by the product of the angular spectrum of the pump beam and the phase matching function. The pump beam is assumed to be a Gaussian beam, which also gives a Gaussian function in the Fourier domain. The phase matching function, on the other hand, is a sinc-function. However, it is often approximated by a Gaussian function to alleviate computations \cite{law}. The latter approximation is quite innoxious in most practical situations where the Rayleigh range of the pump beam is much larger than the length of the nonlinear crystal. In this thin-crystal limit, the phase matching function is effectively evaluated at its origin where it is equal to 1. However, in this limit, one loses the ability to normalize the state. Therefore, we'll retain the Gaussian approximated phase matching function up until a point where it is convenient to apply the limit.

Under these conditions, the SPDC state is given by
\begin{eqnarray}
\psi_{\rm spdc}({\bf a}_1,{\bf a}_2) & = & 2\pi w_p^2\sqrt{2\beta} \exp\left(-\pi^2 w_p^2 |{\bf a}_1+{\bf a}_2|^2 \right) \nonumber \\
& & \times \exp\left(-\frac{1}{2}\pi^2 w_p^2 |{\bf a}_1-{\bf a}_2|^2 \beta \right) ,
\label{spdctoes}
\end{eqnarray}
where ${\bf a}_1$ and ${\bf a}_2$ are the spatial frequency vectors associated with the two respective photons, and
\begin{equation}
\beta = \frac{n_o L}{z_R} = \frac{n_o L\lambda}{\pi w_p^2} ,
\label{betadef}
\end{equation}
is the ratio of the crystal length $L$ (times the ordinary refractive index of the nonlinear crystal $n_o$) to the Rayleigh range. In the thin crystal limit $\beta\rightarrow 0$. However, one needs to remove the factor of $\beta$ from the normalization constant before setting $\beta$ to zero, to avoid setting the whole expression to zero.

The input density matrix in this case is given by
\begin{equation}
R_0({\bf a}_1,{\bf a}_2,{\bf a}_3,{\bf a}_4) = \psi_{\rm spdc}({\bf a}_1,{\bf a}_3) \psi_{\rm spdc}^*({\bf a}_2,{\bf a}_4) .
\label{insdef}
\end{equation}
The complex conjugation has no effect, because the SPDC state is real-valued. We substitute Eq.~(\ref{insdef}) into Eqs.~(\ref{koropl}) and (\ref{onkopl}) and evaluate the integrations over the auxiliary variables of the respective expressions. Then we apply the thin-crystal limit. The results are
\begin{widetext}
\begin{eqnarray}
R({\bf a}_1,{\bf a}_2,{\bf a}_3,{\bf a}_4,t) & = & \frac{8 \pi^2 w_p^4}{N_0} \exp \left\{ -\frac{\pi^2 w_p^2}{N_0}
\left[ \frac{N_1}{6}(|{\bf a}_1|^2+|{\bf a}_3|^2)+\frac{N_1^*}{6}(|{\bf a}_2|^2+|{\bf a}_4|^2) - 2 N_2 ({\bf a}_1\cdot{\bf a}_3) - 2 N_2^* ({\bf a}_2\cdot{\bf a}_4) \right. \right. \nonumber \\
& & \left. \left. - \frac{2 {\cal K} t N_3}{3} ({\bf a}_1\cdot{\bf a}_2+{\bf a}_3\cdot{\bf a}_4) + 4 {\cal K} t N_4 ({\bf a}_1\cdot{\bf a}_4+{\bf a}_3\cdot{\bf a}_2) \right] \right\} ,
\end{eqnarray}
where
\begin{equation}
\begin{aligned}
N_0 & = 8 {\cal K} t + 1 \\
N_1 & = 2 (10 {\cal K}^2 t^4 + 2 {\cal K} t^3 + 12 {\cal K} t + 3) - i 6 t (6 {\cal K} t + 1) \\
N_2 & = ( 2 {\cal K}^2 t^4 - 4 {\cal K} t - 1) - i 2 {\cal K} t^2 \\
N_3 & = 2 (5 {\cal K} t^3 + t^2 + 6) \\
N_4 & = ({\cal K} t^3 - 2) ,
\end{aligned}
\end{equation}
for the case without correlations, and
\begin{eqnarray}
R({\bf a}_1,{\bf a}_2,{\bf a}_3,{\bf a}_4,t) & = & \frac{8 \pi^2 w_p^4}{H_0} \exp \left\{  - \frac{\pi^2 w_p^2}{ H_0} \left[ \frac{H_1}{6} (|{\bf a}_1|^2 + |{\bf a}_3|^2) + \frac{H_1^*}{6} (|{\bf a}_2|^2 + |{\bf a}_4|^2) \right. \right. \nonumber \\
& & \left. \left. + \frac{H_2}{3} ({\bf a}_1\cdot{\bf a}_3) + \frac{H_2^*}{3} ({\bf a}_2\cdot{\bf a}_4) - \frac{4 {\cal K} t H_3}{3} ({\bf a}_1 + {\bf a}_3)\cdot({\bf a}_2 + {\bf a}_4) \right] \right\} ,
\end{eqnarray}
where
\begin{equation}
\begin{aligned}
H_0 & = 16 {\cal K} t + 1 \\
H_1 & = 2 (8 {\cal K}^2 t^4 + 2 {\cal K} t^3 + 24 {\cal K} t + 3) - i 6 t (12 {\cal K} t + 1) \\
H_2 & = 2 (8 {\cal K}^2 t^4 + 2 {\cal K} t^3 + 24 {\cal K} t + 3) + i 24 {\cal K} t^2 \\
H_3 & = (4 {\cal K} t^3 + t^2 + 12) ,
\end{aligned}
\end{equation}
for the case with correlations.
\end{widetext}

These expressions represent the quantum states, expressed as density matrices in the plane wave basis, for the photon pairs propagating through different uncorrelated media or together through the same medium.

\begin{figure*}[th]
\includegraphics{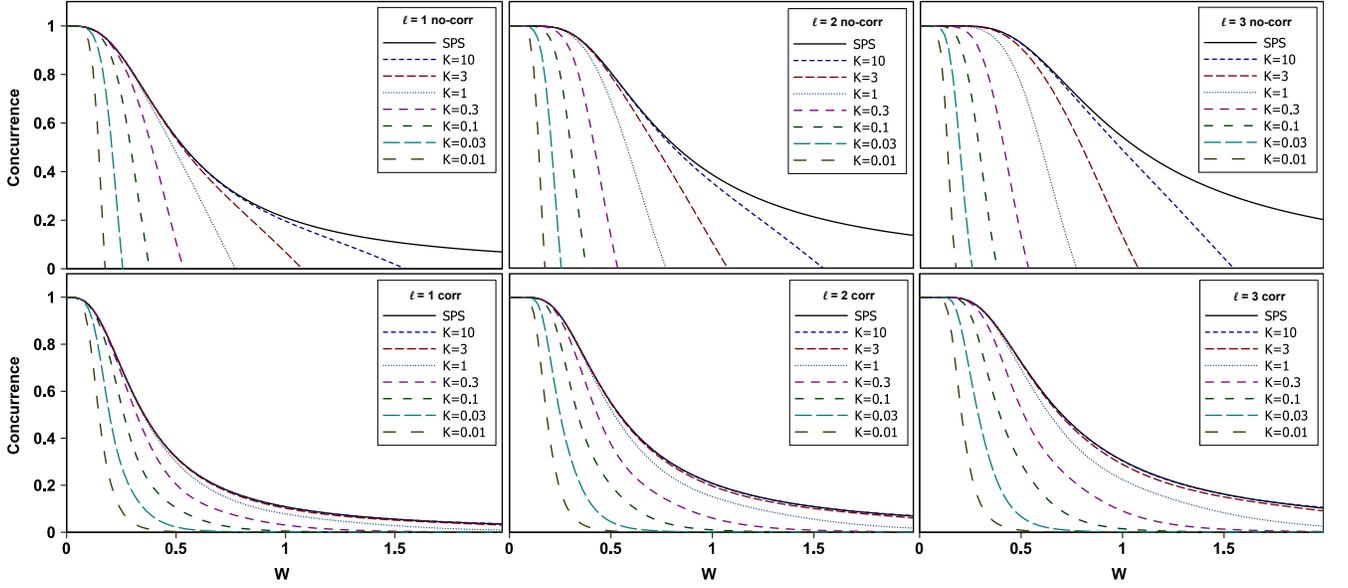}
\caption{Comparison of concurrence curves for the photon pairs propagating through uncorrelated media (top row) or through correlated medium (bottom row).}
\label{sesvol}
\end{figure*}

Next, we consider how much entanglement remains in these states when we project them onto a qubit subspace for each of the two photons. The basis for the qubit subspace is composed of two Laguerre-Gaussian (LG) modes, which are OAM eigenstates. Three different sets of basis functions are considered. All of them consist of two modes with a zero radial index $p=0$ and with the same magnitude in the azimuthal index $|\ell|=q$, where $q=1,2,3$ for the three sets, respectively.

To perform the projection, we compute the overlap between the density matrices for the states and the LG modes, but instead of using the expressions for these respective LG modes, we use a generating function. For $p=0$, the generating function for the angular spectra of the LG modes is given by \cite{ipe,pindex,noncol,inprod}
\begin{eqnarray}
{\cal G}_{\pm}({\bf a},t;\mu) & = & \pi w_0 \exp\left[i \pi w_0(a_x\pm i a_y)\mu \right. \nonumber \\
& & \left. -\pi^2 w_0^2 (a_x^2+a_y^2)(1-i t) \right] ,
\label{genlg}
\end{eqnarray}
where $\mu$ is the generating parameter for the azimuthal index. The sign in the expressions is given by the sign of the azimuthal index. The angular spectrum of a particular LG mode (with $p=0$) is obtained by
\begin{equation}
M_{\rm LG}^{p,\ell}({\bf a},t) = {\cal N}_{\rm LG} \left[ \partial_{\mu}^{|\ell|} {\cal G}({\bf a},t;\mu) \right]_{\mu=0} ,
\label{simodes}
\end{equation}
where
\begin{equation}
{\cal N}_{\rm LG} = \left( \frac{2^{1+|\ell|} }{\pi |\ell|!} \right)^{1/2}
\label{lgn}
\end{equation}
is the modal normalization constant.

After evaluating the overlap integral
\begin{eqnarray}
{\cal H} & = & \int {\cal M}({\bf a}_1,{\bf a}_2,{\bf a}_3,{\bf a}_4,t) {\cal G}_{\pm}^*({\bf a}_1,t;\mu_1) {\cal G}_{\pm}({\bf a}_2,t;\mu_2) \nonumber \\
& & \times {\cal G}_{\pm}^*({\bf a}_3,t;\mu_3) {\cal G}_{\pm}({\bf a}_4,t;\mu_4)\ {\rm d}^2 a_1\ {\rm d}^2 a_2\ {\rm d}^2 a_3\ {\rm d}^2 a_4 , \nonumber \\
\end{eqnarray}
we obtain a generating function for the elements of the density matrix in the projected subspace. Particular elements are computed from this generating function by performing the process in Eq.~(\ref{simodes}) for each of the four generating parameters $\{\mu_1,\mu_2,\mu_3,\mu_4\}$. Note that, since the result is a projection of the full density matrix, it would not be normalized. One needs to divide the projected density matrix by its trace before investigating its entanglement.

The entanglement is quantified by the concurrence \cite{wootters}, which is computed from the eigenvalues of the matrix
\begin{equation}
R = \rho(\sigma_y\otimes\sigma_y)\rho^*(\sigma_y\otimes\sigma_y) ,
\label{rmat}
\end{equation}
where $\rho$ is the density matrix and $\sigma_y$ is the Pauli $y$-matrix. If $\lambda_1$ is the largest eigenvalue, the concurrence is given by
\begin{equation}
{\cal C}\{\rho\} = \max\left\{\sqrt{\lambda_1}-\sqrt{\lambda_2}-\sqrt{\lambda_3}-\sqrt{\lambda_4},0\right\} .
\label{conc}
\end{equation}

The curves for the concurrence as a function of ${\cal W}$ are shown in Fig.~\ref{sesvol}. All graphs in the figure are plotted over the range $0<{\cal W}<2$. The top three graphs represent the cases where the photons propagate through different uncorrelated media. They are produced using qubits composed of LG modes with $|\ell|=1$, $2$ and $3$, respectively. Each graph contains several curves for different values of the dimensionless turbulence strength ${\cal K}$, including the single phase screen (SPS) case, which is obtained by taking the weak scintillation limit, explained in Sec. \ref{weakscint}.

The bottom three graphs in Fig.~\ref{sesvol} represent the cases where both photons propagate through the same (correlated) medium. They are produced using qubits composed of the same LG modes, $|\ell|=1$, $2$ and $3$, respectively, as in the top row and they also contain curves for the same values of the dimensionless turbulence strength ${\cal K}$, including the SPS case.

By comparing the correlated cases with their uncorrelated counterparts, one observes two main differences. The first observation is that there is a scaling of the horizontal dependence. This can best be seen by considering the SPS curves. When one applies the weak scintillation limit to the concurrence curves for the uncorrelated case, the results become relatively simple expressions:
\begin{eqnarray}
{\cal C}_1 & = & \frac{\chi+1}{\chi^2+\chi+1} \label{conc1} \\
{\cal C}_2 & = & \frac{2(\chi+1)(3\chi^2 + 2\chi + 2)}{3\chi^4 + 6\chi^3 + 10\chi^2 + 8\chi + 4} \label{conc2} \\
{\cal C}_3 & = & \frac{(\chi+1)(15\chi^4+24\chi^3+32\chi^2+16\chi+8)}{5\chi^6+15\chi^5+39\chi^4+56\chi^3+48\chi^2+24\chi+8} , \nonumber \\ \label{conc3}
\end{eqnarray}
where
\begin{equation}
\chi = 0.456 {\cal W}^{5/3} .
\label{chidefu}
\end{equation}
If we perform the same weak scintillation limit on the results for the correlated case, we obtain the same three expressions in Eqs.~(\ref{conc1}-\ref{conc3}), but the definition of $\chi$ differs by a factor of 2:
\begin{equation}
\chi = 0.912 {\cal W}^{5/3} .
\label{chidefc}
\end{equation}
This causes a scaling on the horizontal axis. Note that, none of the SPS curves reaches zero at a finite value of ${\cal W}$. This observation differs from the results obtained previously \cite{notrunc}, because the previous results considered a different initial state (a Bell state instead of the SPDC state).

For the other curves (those with other values of ${\cal K}$), there is another difference in addition to this horizontal scaling. While the general concurrence curves becomes zero at a finite value of ${\cal W}$ in the uncorrelated case, those for the correlated case only approach zero asymptotically for increasing ${\cal W}$. This implies that a biphoton propagating through the same medium somehow avoids the entanglement sudden death that is found in the case where the two photons propagate through different uncorrelated media.

\section{Conclusions}
\label{concl}

We derived an evolution equation for a biphoton state propagating through the same turbulent medium, taking into account the fact that such a situation gives rise to correlations between the media seen by the two photons. The derivation follows the infinitesimal propagation approach that gives an equation that is valid under all scintillation conditions and not only under weak scintillation conditions. Throughout the analysis the plane wave basis is used, giving closed form expressions in terms of integrals over the plane wave basis, thus avoiding the truncation problems that can occur for discrete bases.

A solution of the evolution equation is obtained under the quadratic structure function approximation. It has the form of a superposition integral that contains the initial density matrix in the plane wave basis and a kernel function, representing the propagation process.

The solution is studied in the case where the initial density matrix is that of a state prepared with spontaneous parametric down-conversion. Results are compared to those for propagation through different uncorrelated media. It is found that while the uncorrelated media give curves that reach zero at a finite scintillation strength, those for correlated media approach zero asymptotically as a function of the scintillation strength.


\end{document}